\documentclass[12pt,notitlepage,a4paper]{article}

 \pdfoutput=1

\usepackage{epsfig}
\usepackage{color,graphicx}
\usepackage{cite}
\usepackage{amssymb,amsmath}

\newcommand{\be}{\begin{equation}}
\newcommand{\ee}{\end{equation}}
\newcommand{\bea}{\begin{eqnarray}}
\newcommand{\eea}{\end{eqnarray}}

\begin{document}

\begin{center}  

\vskip 2cm 

\centerline{\Large {\bf Embedding massive flavor in ABJM}}
\vskip 1cm

\renewcommand{\thefootnote}{\fnsymbol{footnote}}

   \centerline{Gabi Zafrir \footnote{gabizaf@technion.ac.il} }

\vskip .5cm
{\small \sl Department of Physics} \\
{\small \sl Technion, Haifa 32000, Israel} 

\end{center}

\vskip 0.3 cm

\setcounter{footnote}{0}
\renewcommand{\thefootnote}{\arabic{footnote}}   
   
      \begin{abstract}
       We add massive fundamental matter to the ABJM model by adding D6-branes wrapped asymptotically over $RP^{3}$. We find two types of solutions at finite temperature, one that enters
        the black hole and one that ends before the black hole. We analyze the behavior of the free energy as a function of temperature, and 
       find that the system 
       exhibits a phase transition between the two types of solutions, similar to what happens in the D3-D7 system. We 
       also analyze the meson spectrum in the model and find several massive scalar modes, again, quite like the D3-D7 system. We end with a calculation of the conductivities in the two phases.
      \end{abstract}

    \section{Introduction}
      
      The ABJM\cite{ABJM} model is a conjectured holographic duality between a three dimensional ${\cal N}=6$ supersymmetric Chern-Simons theory with gauge group $U(N)_{k} \times U(N)_{-k}$ and M-theory in an
      $AdS_{4}\times S^{7}/Z_{k}$ background or type IIA string theory on $AdS_{4}\times CP^{3}$. Holographic dualities have been 
      useful in understanding strongly interacting systems. Most of the applications have been to four dimensional systems, such as N=4 SYM. The ABJM model 
      provides an opportunity to study strongly interacting three dimensional systems, which might be applicable to phenomena in condensed matter physics. It is therefore important to extend the model so that it can represent more realistic systems containing matter degrees of freedom.
      
The basic ABJM model contains gauge fields and several scalar and fermionic fields in the bi-fundamental representation. There is no field which is solely in the fundamental of only one of the groups.   

It is possible to extend the model to include fundamental matter by adding D6-branes 
that wrap an $RP^3$ subspace of $CP^3$ in the Type IIA description \cite{HK,GJ,HLT,AEMOT,FT}.
In this paper we will study D6-brane embeddings corresponding to massive matter and explore
the behavior of the system at finite temperature. The D6-brane is treated as a probe in the
AdS-Schwarzschild Type IIA background.
Massive embeddings have been considered previously at zero temperature by Jensen\cite{Jensen}
as a model for a BKT type transition in a background magnetic field. I will show that
at finite temperature there are two kinds of embeddings, corresponding to two possible
phases of the dual field theory. 
In a black-hole (BH) embedding the D6-brane intersects the horizon,
and this corresponds to an ungapped phase in the field theory.
In a Minkowski (MN) embedding the D6-brane ends outside the black-hole horizon,
and this is dual to a gapped phase.
The MN embedding is preferred at low temperature. As the temperature increases there is a
first-order phase transition to a BH embedding.

The behavior of the ABJM-D6 system is qualitatively similar to the D3-D7 system studied in \cite{BEEZI,GSUM,MMT},
and can be viewed as the three-dimensional version of that system.

 The original motivation was to study this system as a possible holographic model for the quantum Hall effect, as suggested in \cite{HLT}. However our calculation of the Hall conductivity in section 6 shows that this system doesn't exhibit quantum Hall phenomena. 

The rest of the paper is organized as follows. In section 2 we will briefly review the ABJM model and explain how to extend it to include matter in the fundamental representation. In section 3 we will derive and solve the equations of motion, and then in section 4 we will exhibit the phase transition. In section 5 we study the meson spectrum. In section 6 we analyze the electrical conductivities. We end in section 7 with some short conclusions.

\section{The geometry of the model}

\subsection{The Type IIA Background}

The ABJM model describes $N$ M2-branes on a $C^4/Z_k$ orbifold singularity.
The low energy field theory is  three dimensional ${\cal N}=6$ supersymmetric Chern-Simons gauge theory with a 
gauge group $U(N)_k\times U(N)_{-k}$. The dynamical fields are four massless complex scalars in the bi-fundamantal
representation and their fermionic partners. At large $N$ this theory has a dual supergravity description corresponding to the
near-horizon background of the M2-branes. For $k\ll N^{1/5}$ this is given by eleven dimensional supergravity in 
$AdS_4\times S^7/Z_k$. For $N^{1/5}\ll k\ll N$ the more appropriate description is in terms of Type IIA supergravity
in $AdS_4\times CP^3$. This will be the regime in which we work.

The $AdS_4 \times CP_3$ metric is given by: 
\begin{eqnarray}
\label{fullmetric}
 ds^2 = \frac{R^3}{k} \left( \frac{1}{4} ds_{AdS_4}^2 + ds_{CP_3}^2 \right)\quad \\
\label{AdSmetric}
\quad ds_{AdS_4}^2 = \frac{r^4}{16} \left( -dt^2 + dx^2 + dy^2 \right) + \frac{4}{r^2} dr^2 \,,
\end{eqnarray}
where R is the radius of the $S^{7}$ in 11d Planck units and is given by\cite{ABJM}:
\be
R^3=2^\frac{5}{2} \pi \sqrt{Nk} \,.
\ee
In a standard parameterization  
the $CP^3$ metric is given by:
\bea
 ds_{CP_3}^2 = d\xi^2 + \cos^2{\xi}\sin^2{\xi}(d\psi + \frac{\cos{\theta_1}}{2} d\phi_1 - \frac{\cos{\theta_2}}{2} d\phi_2 )^2 \nonumber  \\ + \frac{\cos^2{\xi}}{4} (d\theta_1^2 + \sin^2\theta_1 d\phi_1^2) + \frac{\sin^2\xi}{4} (d\theta_2^2 + \sin^2\theta_2 d\phi_2^2) 
\eea
where the ranges of the angles are as follows:
$$0 \leq \xi < \pi/2\quad ,\quad 0 \leq \psi < 2\pi\quad ,\quad0 \leq \theta_i < \pi\quad ,\quad0 \leq \phi_i < 2\pi\quad .$$ 
The dilaton and RR fields are given by\cite{ABJM}:
    
   \be
    e^{2\phi}= \frac{R^{3}}{k^{3}}  
   \ee
   
   \be
    F_{2}=kJ  
   \ee
   
   \be
     F_{4}=\frac{3}{8}R^{3}\epsilon_{4}=\frac{3}{8}R^{3} \left(\frac{r}{2} \right)^{5} dt \wedge dx_{1} \wedge dx_{2}\wedge dr   
   \ee
   where $J$ is the Kahler form of $CP^{3}$, and is given explicitly by

   \bea
   J=-\cos{\xi}\sin{\xi}d\xi\wedge(2d\psi+\cos{\theta_1}d\phi_1-\cos{\theta_2}d\phi_2)  
  \nonumber \\ - \frac{1}{2}\cos^2{\xi}\sin{\theta_1}d\theta_1 \wedge d\phi_1 -    \frac{1}{2}\sin^2{\xi}\sin{\theta_2}d\theta_2 \wedge d\phi_2.
   \eea

There is a generalization of the ABJM model (known as the ABJ model), which allows for a diffrence, $l$, between the ranks of the two groups \cite{ABJ}. In this case the Type IIA dual has a non-zero B-field given by:

\be
B_2 =\frac{l}{k}J.  \label{eq:bt}
\ee 

As we are interested in the behavior of the system at finite temperature, we will need the metric when temperature is taken into account.
   At finite temperature the relevant background contains a black hole, and the metric becomes:

\bea
 ds_{AdS_4}^2 = \frac{r^4}{16} \left( -\left(1-\frac{r^6_H}{r^6}\right)dt^2 + dx^2 + dy^2 \right) + \frac{4}{\left(1-\frac{r^6_H}{r^6}\right)r^2} dr^2 ,
\eea
where $r_H$ is the horizon radius given by: $r^2_H=\frac{16 \pi T }{3}$.

\subsection{Adding flavor}
  
  Now we shall describe the embedding.
First, we should mention that once we reduce from M-theory to type IIA the M2 branes become D2 branes living on the t ,$x_{1}$ , $x_{2}$ coordinates of $AdS_4$. We want to add fundamental matter by adding a single probe D6-brane, so that the strings between it and the D2 branes will give us fundamental fields. The distance between these branes will determine the mass of these fields. 
  
Our starting point is the massless embedding \cite{HK,GJ,HLT,AEMOT,FT}, in which a D6-brane is embedded on all of $AdS_{4}$ and on an $RP^{3}$ subspace of $CP^{3}$, which is the Lagrangian submanifold. The $RP^3\subset CP^3$ that the D6-brane wraps is given by:

$$ \theta_1 = \theta_2 = \theta , \; \; \phi_1 = -\phi_2 = \phi, \; \;  \xi = \frac{\pi}{4}. $$
 This embedding preserves $\cal{N}=$ 3 supersymmetry. Since the D6-brane shares all the coordinates of the D2-branes, the matter fields live in all 3 dimensions of the dual field theory. This gives the field theory with the quiver diagram shown in figure 1.

\begin{figure}[h]
\center
\includegraphics[width=0.8\textwidth]{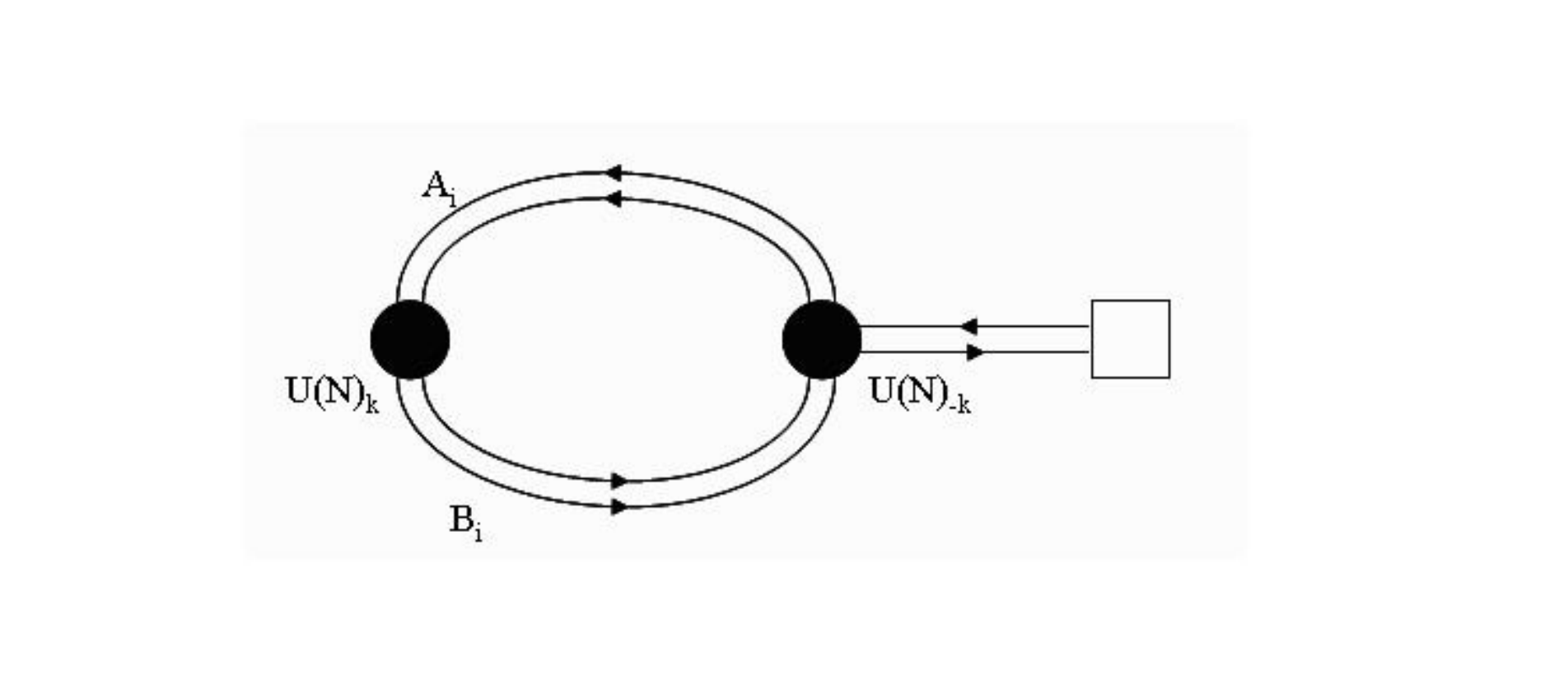} 
\caption{Quiver diagram for the field theory with flavor.}
\label{fig:massr0}
\end{figure}

We will consider a more general embedding given by $ \xi = \xi (r) $, with $\xi (r\rightarrow\infty)=\frac{\pi}{4}$.   
This corresponds to a mass deformation for the matter fields. In order to compute the mass we need to
undo the large $N$ limit and measure the D2-D6 separation. This is not easy to do directly in the Type IIA
description, since the seven-dimensional geometry transverse to the D2-branes at finite $N$ is not
well-understood. However, we can start with the flat space Type IIB brane configuration of \cite{ABJM}
and work our way to the Type IIA picture. Recall that the Type IIB configuration consists of an NS5-brane,
a $(1,k)$ 5-brane and D3-branes arranged as follows: 

\begin{center}
\begin{tabular}{ccccccccccc}
& 0 & 1 & 2 & 3 & 4 & 5 & 6 & 7 & 8 & 9 \\
NS5 & $\bullet$ & $\bullet$ & $\bullet$ & $\bullet$ & $\bullet$ & $\bullet$ & & &  \\
$(1,k)5$ & $\bullet$ & $\bullet$ & $\bullet$ & $\cos\theta$ & $\cos\theta$ &$\cos\theta$ &
& $\sin\theta$ & $\sin\theta$ & $\sin\theta$ \\
D3 & $\bullet$ & $\bullet$ & $\bullet$ & & & & $\bullet$ & & & 
\end{tabular}
\end{center}

In this picture fundamental matter is incorporated by adding a D5-brane along $(012789)$ \cite{HK,GJ,HLT,AEMOT,FT}.
Since the background spacetime is flat in this description, the mass of the fundamental is given
simply by the D3-D5 separation along $(345)$.
The Type IIA supergravity dual is obtained by first T-dualizing along $x^6$, then lifting to M theory, taking the near-horizon limit, and
finally reducing along the Hopf fiber to Type IIA string theory. The transformation of coordinates is given explicitly by \cite{HLT}:

\be
   x^6=\psi
   \ee
   
   \be
   \vec{x'_1}=\vec{x_1}=r^4\cos^2\xi(\cos\theta_1,\sin\theta_1\cos\phi_1,\sin\theta_1\sin\phi_1)
     \ee
     
     \be \vec{x'_2}=\vec{x_1}+k\vec{x_2}=r^4\sin^2\xi(\cos\theta_2,\sin\theta_2\cos\phi_2,\sin\theta_2\sin\phi_2)     
   \ee
where $\vec{x}_1=(x^7,x^8,x^9)$ and $\vec{x}_2=(x^3,x^4,x^5)$.
Furthermore, since the above relation involves T-duality transverse to NS5-branes, there is a non-trivial metric given by \cite{ABJM}:

\bea
   ds^2=\frac{d\vec{x'_1}\cdot d\vec{x'_1}}{2|\vec{x'_1}|}+\frac{d\vec{x'_2}\cdot d\vec{x'_2}}{2|\vec{x'_2}|}.
   \eea   
In this description the mass is given by the proper length of a (geodesic) path of fixed $\vec{x}_1$, {\em i.e.},

\bea
s = \int \frac{d|\vec{x}'_2|}{\sqrt{2|\vec{x}'_2|}} = \sqrt{2} (\sqrt{|\vec{x}_1 + k\vec{x}_2|} - \sqrt{|\vec{x}_1|}) \\ \nonumber
= \sqrt{2} r^2 (\sin\xi - \cos\xi) = 2r^2 \sin(\xi-\frac{\pi}{4})
\eea
   The mass is given by    
   \be
   M=\frac{s(r\rightarrow\infty)}{2\pi}. \label{eq:MF}
   \ee   

This embedding preserves some of the symmetries of $CP^3$. The asymptotic space, $RP^3$, has an $SO(4)$ symmetry which is locally equivalent to $SO(3)\times SO(3)$. This embedding preserves one of the $SO(3)$'s (which corresponds to the R-symmetry in the field theory) and breaks the other to $U(1)$ (which is a global symmetry in the field theory).
   
\section{Derivation and solution of the equation of motion}
   
   Next we determine the behavior of $ \xi (r) $ from the equation of motion of the brane. The relevant part of the D6-brane action is given by $S_{DBI}+S_{CS}$, where:
       
   \be
    S_{DBI}= -\mu_{6} \int e^{-\Phi}\sqrt{-det(G_{ab})}d^{7}\sigma 
    \ee
   
   \be
    S_{CS} = -\mu_6\int P[ C_7].
    \ee
    
    The 7-form potential can be evaluated by dualizing $ F_{2}$ to get $ F_{8}$ and then integrating to get $ C_{7} $. The process is not difficult but quite tedious so we will simply write the result:
   
   \be
   dC_{7}=-\frac{R^{9}r^{5}}{2^{11}k^{2}}J \wedge J \wedge dr \wedge dt \wedge dx_{1} \wedge dx_{2}.
   \ee        
In a specific gauge choice:

\be
 C_{7}=-\frac{R^{9}r^{6}}{3\cdot2^{12}k^{2}}J \wedge J \wedge dt \wedge dx_{1} \wedge dx_{2}, \label{eq:gauge}
\ee
giving the following pullback to the brane worldvolume:

\be
    P[C_{7}]=-\frac{R^{9}r^{6}}{3\cdot2^{12}k^{2}} \cos 2\xi \sin 2\xi \sin\theta  \xi' dr \wedge d\theta \wedge d\psi \wedge d\phi \wedge dt \wedge dx_{1} \wedge dx_{2}.
    \ee
The induced metric on the D6-brane is:

   \bea
    ds_{D6}^2 = \frac{R^3}{4k} \left[\frac{r^4}{16} \left( -\left(1-\frac{r^6_H}{r^{6}}\right)dt^2 + dx^2 + dy^2 \right) + \left(\frac{4}{r^2(1-\frac{r^6_H}{r^{6}})} +  4\xi'^2\right)dr^2 \right. \nonumber \\ \left. + d\theta^2 + \sin^2\theta d\phi^2 + \sin^2 2\xi (d\psi + \cos\theta d\phi)^2 \right]. \label{eq:xiS}
    \eea

    The D6-brane worldvolume action then reduces to:
    
    \be
     S = \frac{R^{9}}{2^{11}k^{2}} \int dr r^{6} \sin(2\xi)\left(\sqrt{r^{-2}+\xi'^2}   - \frac{1}{3}\cos(2\xi) \xi'\right) \ \label{eq:xiL} 
    \ee
    and the EOM for $ \xi $ can be easily derived:
           
    \bea
  \partial_r \left[ \frac{r^7 \sin(2\xi) \xi' (1-\frac{r^6_H}{r^{6}})}{\sqrt{1+\xi'^{2} r^{2} (1-\frac{r^6_H}{r^{6}})}}\right] = 2r^{5}  \cos(2\xi) \left(\sqrt{1+\xi'^{2} r^{2} \left(1-\frac{r^6_H}{r^{6}}\right)}  + \sin(2\xi) \right). \nonumber \\  \label{eq:xiEOM}
 \eea
       Note the trivial solution $\xi=\pi/4$, corresponding to the massless embedding. 
    The asymptotic form of the general solution is found to be: 
       
\be
\xi \sim \frac{\pi}{4}+\pi M r^{\alpha_+} + c r^{\alpha_-}, \label{eq:xi}
\ee
with $\alpha_-=-4$ and $\alpha_+=-2$.\footnote{As in other AdS-CFT examples, this is related to the dimension of the dual operator. Note however that our radial coordinate is different 
than the one related to energy scaling. The relation being 
$r_{Energy}=(r_{Here})^2$, so the dimensions of the field are 1, 2 consistent with $mass^2=-2$ and with the results of previous works}
  The leading term is proportional to the mass, where the constant can be calculated using the mass formula (\ref{eq:MF}). The normalizable mode, in accordance with the state-operator correspondence, is related to the expectation value of the dual operator. Therefore, this term corresponds to the flavor condensate $c$.
                          
     From (\ref{eq:xiS}) one can see that the brane can end at some $r_{0}>r_{H}$, where $\xi(r_{0})=0$ or $ \frac{\pi}{2}$. Indeed, as was mentioned, there are two possible cases: either the brane reaches the horizon or it ends before it. The behavior of the solution in each case is different.
           
    The equation of motion can be solved numerically for both types of solutions. In the black hole embedding, one starts from the horizon with some initial value $\xi(r_H)$, which determines the mass. The derivative at the horizon $\xi'(r_H)$ is fixed by the EOM:
    
    \be
    \xi'(r=r_H)=\frac{\cot{2\xi_0}(1+\sin{2\xi_0})}{3r_H}.
    \ee
    
     In the Minkowski embedding one starts at some $r=r_0>r_H$, which now determines the mass, and $\xi'(r_0)\rightarrow\infty$ (in practice we use vanishing derivative of $r(\xi)$ ). Once the solution is determined one calculates the mass from (\ref{eq:xi}). The results we found are shown in figure 2.

\begin{figure}[h]
\center
\includegraphics[width=0.45\textwidth]{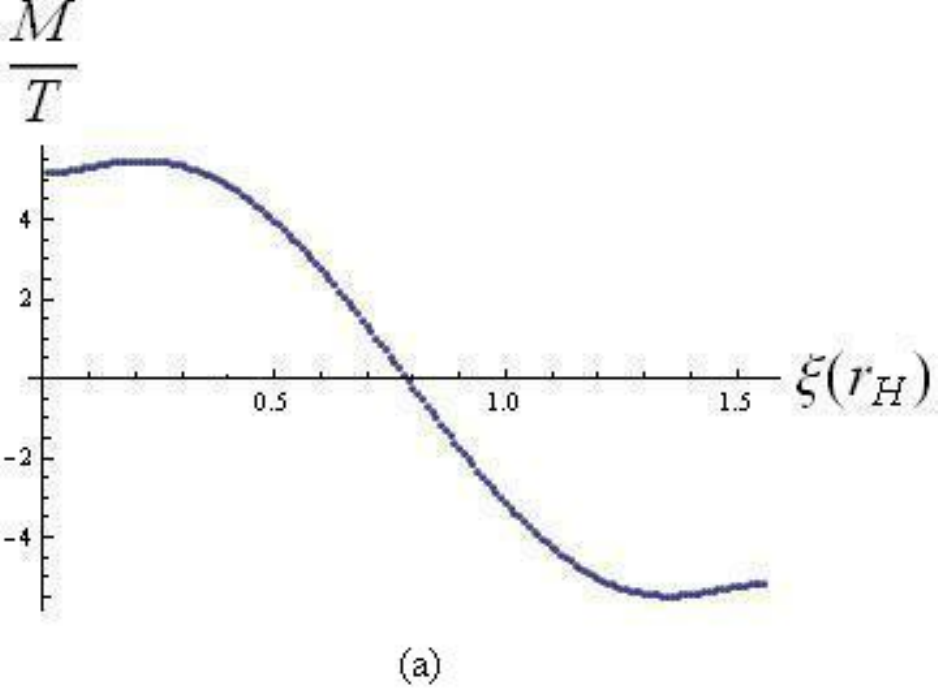}
\hspace{0.5cm}
\includegraphics[width=0.45\textwidth]{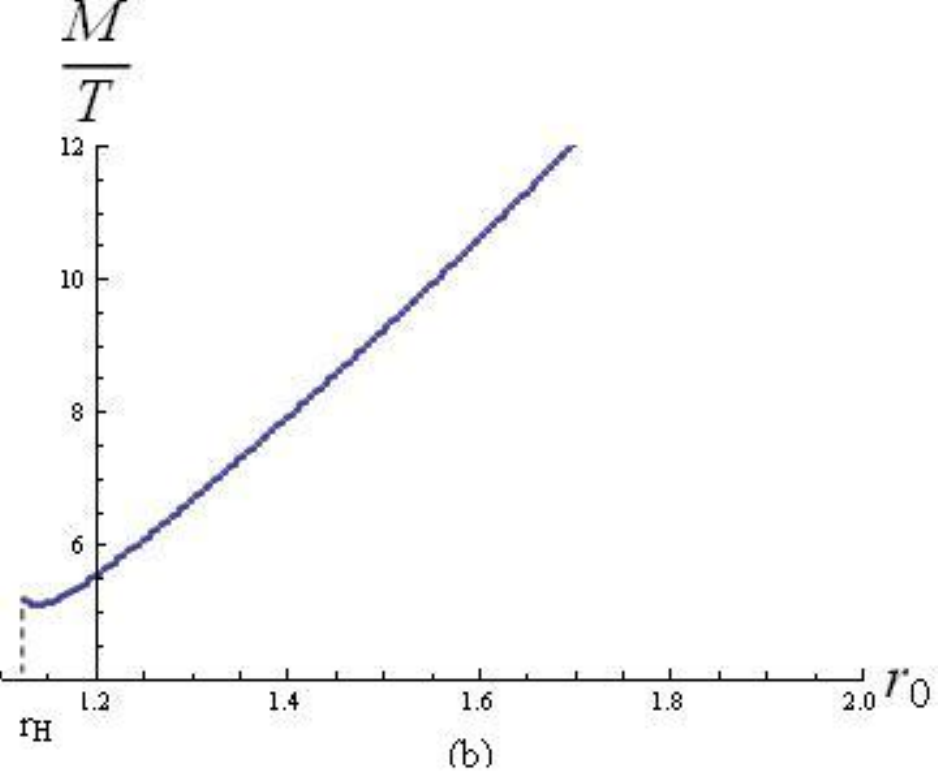} 
\caption{$\frac{Mass}{T}$ as a function of initial parameter at fixed temperature: (a) BH solution as function of the initial angle. (b) MN solution as function of the initial $r$. 
($r^6_H=2$)}
\label{fig:massr0}
\end{figure}

    Since the equations are invariant under $\xi\rightarrow \frac{\pi}{2} -\xi$, every embedding (MN or BH) has a ``mirror" embedding, but with the opposite sign of the mass. This is apparent in figure 1a, which shows the BH solutions.  
     Figure 1b shows MN solutions with $ \xi(r_0) = 0 $. There are corresponding ones for $ \xi(r_0) = \frac{\pi}{2} $. 
     
     \section{Phases of the model}
     
     Looking at figure 2, we see that when $ M >> T $ there is only one solution, which is an MN embedding far from the horizon. As the temperature increases two BH solutions and another MN solution appear. When the temperature is further raised the MN solutions disappear and we are left with only a single black hole solution.
    This suggests that there should be a phase transition between these two states at some temperature. We next set out to confirm this. One way to do this is to examine the $c$ vs. $\frac{M}{T}$ curve which we have calculated. The calculation is quite similar to the evaluation of the mass range, but now we used the value of $\xi$ and its derivative at large $r$ to calculate $M$ and $c$ by solving the following two equations:
    
    \be
     \xi(r\rightarrow\infty) \approx \frac{\pi}{4}+\frac{\pi M} {r^2} + \frac{c} {r^4} 
    \ee
    
    \be
    \xi'(r\rightarrow\infty) \approx -2\frac{\pi M} {r^3} - 4\frac{c} {r^5}.
    \ee  
    The results are shown in figure 3.
    
    First, one notices again the $\xi\rightarrow \frac{\pi}{2} -\xi$ reflection symmetry where on the right are embeddings with $\xi<\frac{\pi}{4}$ while on the left $\xi>\frac{\pi}{4}$. More importantly, there is a region where the graph is not single-valued, this is typical in cases of a first order phase transition, and confirms that this also happens in this case. Furthermore, one can use the Maxwell construction to calculate the critical value of $\frac{M}{T}$. 
    
    \begin{figure}[h]
\center
\includegraphics[width=0.5\textwidth, height=0.25\textheight]{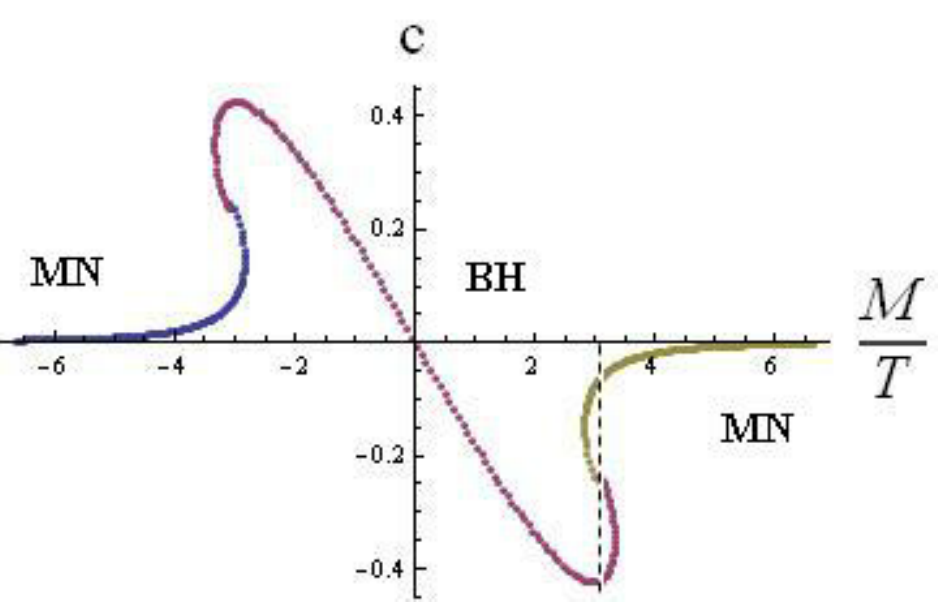}
\caption{c vs $\frac{M}{T}$ at constant temperature: blue and green are MN embeddings while the purple are the BH embeddings. The Maxwell construction is illustrated so one can infer the critical value of $\frac{M}{T}$.}
\label{fig:cvsm}
\end{figure}

     Another way to show the phase transition is to evaluate the free energy of each solution, for a fixed mass, as a function of temperature. The free energy of a solution is proportional to its Euclidean action:
   
\bea
   F\propto S_E.
\eea    
 This diverges at large $r$. Therefore, we must regularize it. We used holographic renormalization, adding the following counterterm:
 
 \be
 S_{counterterm} = -\frac{k}{6}\int_{bound.} d^3 x \sqrt{-\gamma} \left(1-2(\xi-\frac{\pi}{4})^2\right),
 \ee
 where $\gamma$ is the induced metric on the UV cutoff. This renders the action finite. However since there is a CS term one must also add a boundary term so that the action will be gauge invariant. Once the gauge choice (\ref{eq:gauge}) is made there remains the gauge symmetry:
 
 \be
 C_7= -\frac{R^{9}r^{6}}{3\cdot2^{11}k^{2}} \cos 2\xi \sin 2\xi \xi' \rightarrow -\frac{R^{9}(r^{6}+C)}{3\cdot2^{11}k^{2}} \cos 2\xi \sin 2\xi  \xi'. \label{eq:Gtran}
 \ee
 Hence, we add the following boundry term:
 
 \be
 S_{boundary} = \frac{R^9}{2^{11}k^2} \int_{bound.} d^3 x \frac{r^6}{12}\sin^2(2\xi),
 \ee
 so that under (\ref{eq:Gtran}) it transforms as:
 
 \be
 \frac{R^9}{2^{11}k^2} \int_{bound.} d^3 x \frac{r^6}{12}\sin^2(2\xi) \rightarrow \frac{R^9}{2^{11}k^2} \int_{bound.} d^3 x \frac{r^6+C}{12}\sin^2(2\xi),
 \ee
 and the action remains invariant. We can now use this gauge symmetry to set $C_7(r=r_H)=0$, and so we have:
 
 \be
 C_7= -\frac{R^{9}r^{6}}{3\cdot2^{11}k^{2}} (1-\frac{r^6_H}{r^6}) \cos 2\xi \sin 2\xi \xi',
 \ee
 
 \be
 S_{boundary} = \frac{R^9}{2^{11}k^2} \int_{bound.} d^3 x \frac{r^6}{12} (1-\frac{r^6_H}{r^6}) \sin^2(2\xi).
 \ee
 
 We then scanned over a range of temperatures, evaluated the solutions with the appropriate mass for each temperature and calculated their free energy. 
     A typical plot of the free energies is shown in figure 4.
       
\begin{figure}[ht]
\center
\includegraphics[width=1\textwidth, height=0.5\textheight]{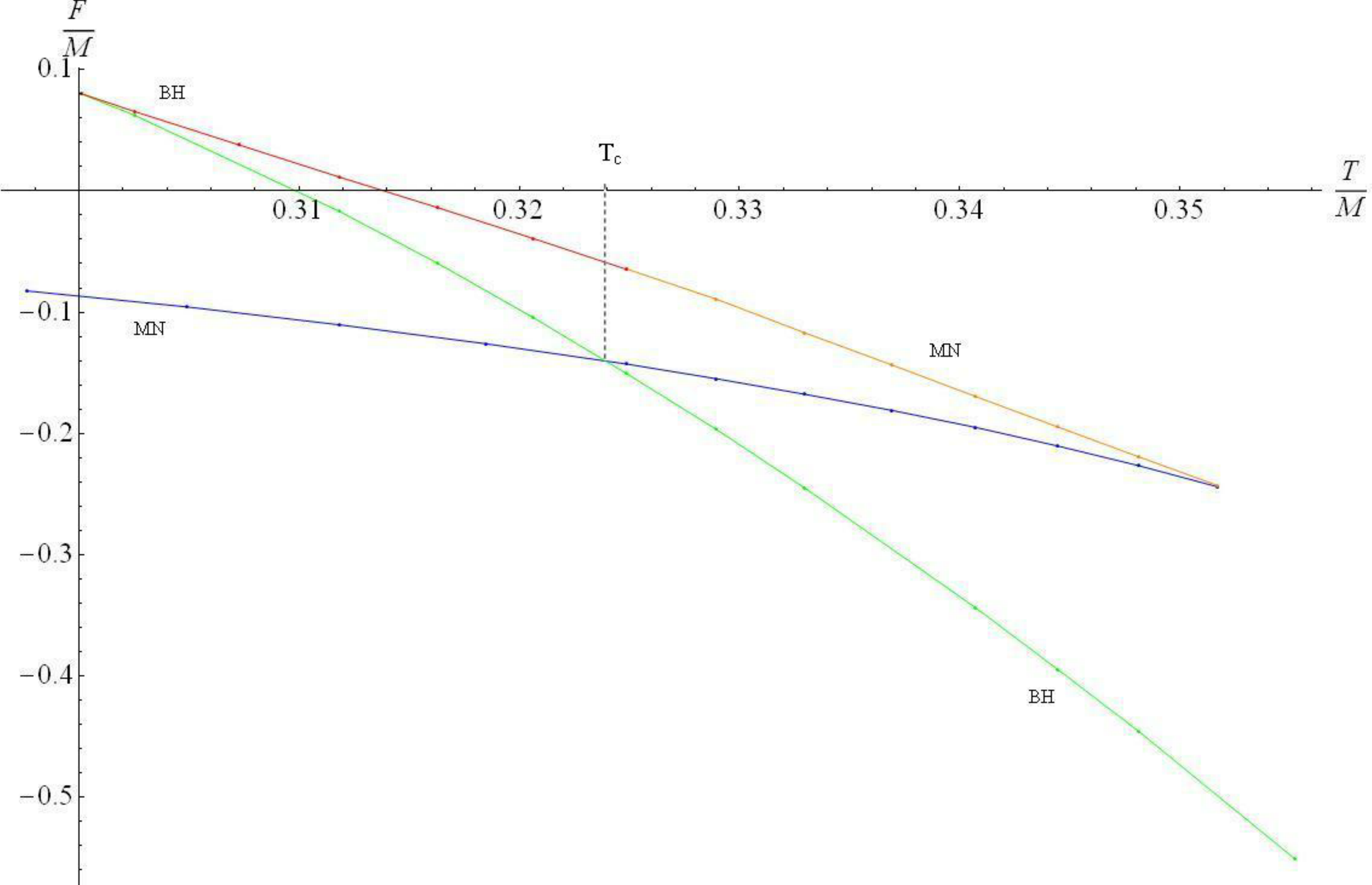}
\caption{Free energy vs temperature: Blue and orange are MN embeddings while red and green are BH embeddings. Both the free energy and the temperature are scaled by the mass of the embedding.}
\label{fig:fvsk}
\end{figure}
    
    As expected, one can see that at low temperatures the MN solution is the only possible one. As the temperature increases, two BH solutions appear, but are unfavored. Then at some temperature one of the BH solutions becomes favored and a phase transition occurs. Also, at some temperature 
    the other BH solution pinches off and becomes an MN solution (one can see in figure 2 that the masses for $\xi(r_H)\approx0$ in (a) and $r_0\approx r_H$ in (b) are nearly identical), 
which eventually merges with the original MN solution. At high enough temperatures there is only the BH solution.
    
    \section{Meson spectrum}
    
    We will now analyze the meson spectrum corresponding to the fluctuations of the probe brane worldvolume fields. There are three scalar fields related to the directions normal to the brane, and there is the gauge field which gives one scalar field and one vector meson. Furthermore, there are various fermionic fields which are the superpartners of the bosonic fields. I will look at only the first three scalar fields, corresponding to the three embedding coordinates. We will use the following notation:
    
    \be
    \xi=\xi_0 + \delta\xi, \hspace{1cm}  \frac{\theta_1-\theta_2}{2}=\alpha, \hspace{1cm}  \frac{\phi_1+\phi_2}{2}=\beta,
    \ee
     where $\xi_0$ is the solution of equation (\ref{eq:xiEOM}), and it is assumed that $\delta\xi, \alpha, 
    \beta << 1$ and that they are independent of the $CP^{3}$ coordinates, but may depend on the $AdS_{4}$ coordinates\footnote{Dependence on the coordinates of $CP^{3}$ will simply give angular momentum internal modes for each meson.}. Next, we need to evaluate the probe brane action, which will now depend on the fields $\delta\xi, \alpha, \beta$. As these fields are very small we can expand the action to the first non-vanishing order (which is quadratic as we expand around a solution to the probe brane's EOM). We will have a system of coupled linear second order differential equations. The eigenvalues of these equations are the masses of the mesons. 
    
    However calculating the full action when all three fields are present is quite tedious. This can be eased by noticing that $\beta$ decouples from the rest. The reason is that the metric is independent of $\phi_1, \phi_2$, and therefore $\beta$ enters the equations only through derivatives. These are either second order and diagonal in $\beta$ (like $(d\beta)^2$), or first order terms but non-diagonal (like $d\beta d\phi$, which will give a term on $AdS_4, \phi$). The diagonal terms are already second order in $\beta$ and therefore cannot couple. The non-diagonal terms always multiply one another (because the original $AdS_4$ metric is diagonal) generating terms which are second order in $\beta$. This makes $\beta$ easier to analyze and so we shall start with this field.  
    
    Therefore we take:
    
    \be
    \xi=\xi_0 , \hspace{1cm} \frac{\theta_1-\theta_2}{2}=0, \hspace{1cm}  \frac{\phi_1+\phi_2}{2}=\beta<< 1.
    \ee
     Expanding the Lagrangian to second order gives:
     
     \bea
   & L[\beta] = L_0 + \Omega r^5 \sin^3{2\xi_0}\left[\frac{r^2 (1-\frac{r^6_H}{r^6}) \beta'^2}{4\sqrt{1+(1-\frac{r^6_H}{r^6})(r\xi'_0)^2}} \right.  \nonumber \\ & \quad  \left.
    -\frac{16}{r^4} \sqrt{1+\left(1-\frac{r^6_H}{r^6}\right)(r\xi'_0)^2} \left(\frac{\dot{\beta}^2}{1-\frac{r^6_H}{r^6}} -(\partial_{x_1} \beta)^2 - (\partial_{x_2} \beta)^2\right)\right].
    \eea
      Where $L_0$ is the zeroth order lagrangian giving the action (\ref{eq:xiL}), and $\Omega$ is some numerical constant.
      The equation of motion for $\beta$ is then:
      
      \bea
     & \partial_r \left(\frac{r^7 \sin^3{2\xi_0} \left(1-\frac{r^6_H}{r^6}\right) }{\sqrt{1+\left(1-\frac{r^6_H}{r^6}\right)(r\xi'_0)^2}} \beta' \right) =  \nonumber \\ & \quad
       64 r \sin^3{2\xi_0} \sqrt{1+\left(1-\frac{r^6_H}{r^6}\right)(r\xi'_0)^2} \left(\frac{\ddot{\beta}}{1-\frac{r^6_H}{r^6}}-\partial^2_{x_1} \beta - \partial^2_{x_2} \beta\right)
      \eea
            
     We solve the equation by separation of variables:
     
     \be
       \beta= \beta(r)e^{-i\omega t}e^{i\textbf{k}\cdot \textbf{x}}. \label{eq:sep}
    \ee
    We then get the following equation for $\beta(r)$:
    
    \bea
     & \partial_r \left(\frac{r^7 \sin^3{2\xi_0} (1-\frac{r^6_H}{r^6}) \beta'}{\sqrt{1+(1-\frac{r^6_H}{r^6})(r\xi'_0)^2}}\right) =  \nonumber \\ & \quad
       64 r \sin^3{2\xi_0} \sqrt{1+\left(1-\frac{r^6_H}{r^6}\right)(r\xi'_0)^2} \left(|\textbf{k}|^2-\frac{\omega^2}{1-\frac{r^6_H}{r^6}}\right)\beta. \label{eq:beta}
      \eea
    We then need to give either $\omega$ or $|\textbf{k}|$ and determine the other such that the solution is normalizable. Since the system at finite temperature is not Lorentz invariant, one must be careful about the definition of the rest mass. We follow\cite{MMT} and define the rest mass as $m=\omega^2$ in the rest frame, that is when $|\textbf{k}|=0$. Therefore we set $|\textbf{k}|=0$ and solve for the values of $\omega$ for which $\beta(r)$ is normalizable.
    
    First we evaluate the asymptotic $r\rightarrow\infty$ dependence, by expanding (\ref{eq:beta}) to the first non-vanishing order in $r^{-1}$, finding:
    
    \be
    \beta(r\rightarrow\infty) \approx A + B r^{-6}.
    \ee
    There should therefore be one normalizable solution and one non normalizable solution. 
     
\begin{figure}[h]
\center
\includegraphics[width=0.7\textwidth, height=0.35\textheight]{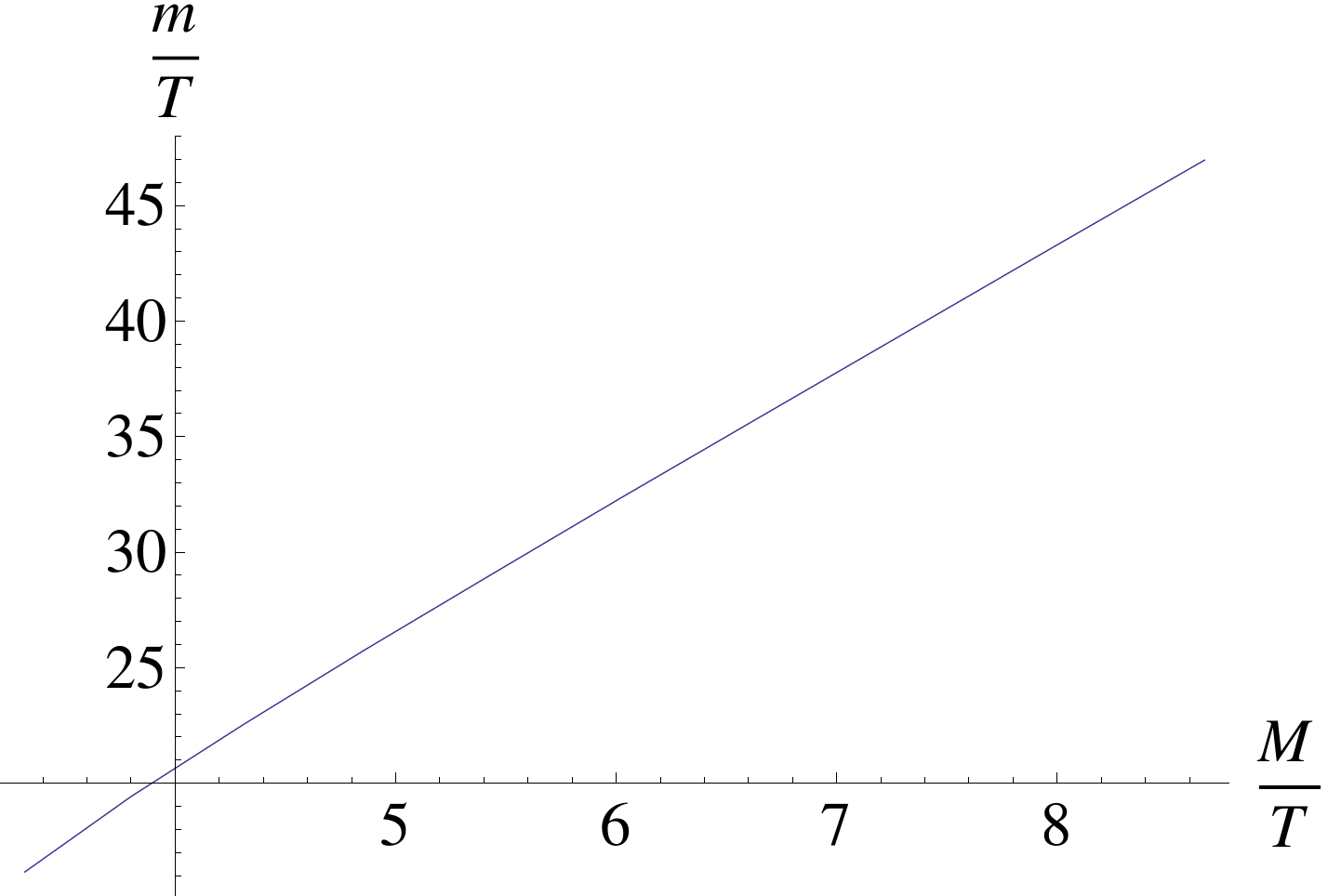}
\caption{The mass of the first mode of $\beta$ as a function of the embedding mass, for MN embeddings. Both masses are scaled by the temperature.}
\label{fig:fvsk}
\end{figure}

    The equation is then solved numerically. First, we shall discuss the MN embedding. 
    What we did in practice is to integrate the equation starting from a small value of $\beta(r)$, with vanishing derivative, near $r=r_0$. The solution then diverges at large r, going to: $\beta(r\rightarrow\infty)\rightarrow\pm\infty$. We scanned over $\omega$ and identified the normalizable mode by the value of $\omega$ where the asymptotic behavier of $\beta$ changes from $\beta(r\rightarrow\infty)\rightarrow+\infty$ to $\beta(r\rightarrow\infty)\rightarrow-\infty$. 
    The result for the lowest mode of $\beta$ is shown in figure 5.

    In the BH embedding there are no normalizable modes, as can be seen by expanding the EOM (\ref{eq:beta}) around $r\approx r_H$ and solving for the behavior of $\beta$ in that region. One finds:
    
    \be
    \beta\approx A\cos{\left(\frac{4\omega}{3}\ln{(r-r_H)}\right)} + B\sin{\left(\frac{4\omega}{3}\ln{(r-r_H)}\right)}.
    \ee
    This is ill-defined as $r\rightarrow r_H$ and therefore the solutions are non-normalizable. This makes sense, since in the BH case, there is nothing preventing the mesons from falling into the horizon and hence deeming them unstable.
    
    Next we will analyze $\xi$ and $\alpha$. Though it isn't obvious a-priori, the two fields actually decouple and so, to simplify matters, we will expand them separately. We will start with $\alpha$ and take (as $\beta$ also decouples):
    
    \be
    \xi=\xi_0 , \hspace{1cm}  \frac{\theta_1-\theta_2}{2}=\alpha, \hspace{1cm}  \frac{\phi_1+\phi_2}{2}=0.
    \ee 
    With this choice the determinant can be broken into two pieces: an AdS + $\theta$ part and a $\psi + \phi$ part. We will start with the $\psi + \phi$ part, which gives:
    
    \bea
    \frac{\sin{2\xi}}{4} \int^{\pi}_{0} \sqrt{\sin^2{\theta}+\alpha \sin{2\theta} \cos{2\xi}+\alpha^2 \cos{2\theta}} d\theta.
    \eea
     Expanding the square root to quadratic order gives:
    
    \bea
    \frac{\sin{2\xi}}{4} \int^{\pi}_{0} \sin{\theta}\left(1+\alpha \cot{\theta} \cos{2\xi}+\frac{\alpha^2 \cos{2\theta}}{2\sin^2{\theta}}\right) d\theta.
    \eea
      
      Note however that the last term gives a divergent contribution. Doing the integral first and then expanding does not work either. The integral can be expressed as an elliptic integral but cannot be expanded around $\alpha=0$.
      There appears to be a problem at $\theta=0, \pi$. This can be understood from the definition of $\theta, \alpha$ and from the range of the angles $\theta_1, \theta_2$. The only way to have $\theta=0$ is if $\theta_1= \theta_2 =0$ and therefore $\alpha=0$, and likewise for $\theta=\pi$. So $\theta=0, \pi$ and $\alpha\neq 0$ is inconsistent and this is the reason for the divergence.
      $\alpha$ must depend on $\theta$ such that $\alpha=0$ at $\theta=0, \pi$.
      This does not change the splitting of the determinant or the $\psi + \phi$ part we calculated. Now we need to calculate the AdS + $\theta$ part, expand it in $\alpha$ and multiply it by the $\psi + \phi$ part. This is quite lengthly and uneventful, and in the end we get:
      
       \bea
       L[\alpha]= \Omega r^5 \sin^3{2\xi_0}\sqrt{1+\left(1-\frac{r^6_H}{r^6}\right)(r\xi'_0)^2}\left[\frac{r^2 \left(1-\frac{r^6_H}{r^6}\right) \alpha'^2 \sin{\theta}}{8(1+\left(1-\frac{r^6_H}{r^6}\right)(r\xi'_0)^2)} \right.  \nonumber \\ \left. - \frac{8\sin{\theta}}{r^4}\left(\frac{\dot{\alpha}^2}{1-\frac{r^6_H}{r^6}} -(\partial_{x_1} \alpha)^2 - (\partial_{x_2} \alpha)^2\right)  +\frac{1}{2} \sin{\theta} (\partial_\theta \alpha)^2 + \frac{\alpha^2}{2\sin{\theta}}\cos{2\theta}\right],
       \eea
      where I have used a dot and a prime for the time and r derivative respectively.  
      We can now derive the EOM for $\alpha$:
      
      \bea
     & \partial_r \left[\frac{r^7 \sin^3{2\xi_0} \left(1-\frac{r^6_H}{r^6}\right) \alpha'}{\sqrt{1+\left(1-\frac{r^6_H}{r^6}\right)(r\xi'_0)^2}}\right] = \sin^3{2\xi_0} \left[ 
       64 r  \sqrt{1+\left(1-\frac{r^6_H}{r^6}\right)(r\xi'_0)^2} \left(\frac{\ddot{\alpha}}{1-\frac{r^6_H}{r^6}}-\partial^2_{x_1} \alpha - \partial^2_{x_2} \alpha \right) \right. \nonumber \\ & \quad \left.
       - 4 r^5  \sqrt{1+\left(1-\frac{r^6_H}{r^6}\right)(r\xi'_0)^2} \frac{\partial_\theta (\sin{\theta} \partial_\theta \alpha)}{\sin{\theta}} + \frac{4\alpha r^5  \cos{2\theta}}{\sin^2{\theta}}\sqrt{1+\left(1-\frac{r^6_H}{r^6}\right)(r\xi'_0)^2}  \hspace{0.01cm} \right].
      \eea
      Proceeding by separation of variables we will take $\alpha=\alpha_A(r,t,x_1,x_2)\alpha_\theta(\theta)$, and after some rearrangement we get:
      
      \bea
      \frac{1}{r^5 \sin^3{2\xi_0} \sqrt{1+\left(1-\frac{r^6_H}{r^6}\right)(r\xi'_0)^2} \alpha_A}\partial_r \left[\frac{r^7 \sin^3{2\xi_0} \left(1-\frac{r^6_H}{r^6}\right) \alpha_A '}{\sqrt{1+\left(1-\frac{r^6_H}{r^6}\right)(r\xi'_0)^2}}\right] \nonumber \\ - \frac{64}{\alpha_A r^4} \left[\frac{\ddot{\alpha}}{1-\frac{r^6_H}{r^6}}-\partial^2_{x_1} \alpha - \partial^2_{x_2} \alpha\right]
      = \frac{4\cos{2\theta}}{\sin^2{\theta}}-\frac{\partial_\theta (\sin{\theta} \partial_\theta \alpha_\theta)}{\sin{\theta}\alpha_\theta}= k^2.
      \eea
      
      We get the following two equations for the two parts:
      
      \bea
     &  \partial_r \left(\frac{r^7 \sin^3{2\xi_0} \left(1-\frac{r^6_H}{r^6}\right) \alpha_A '}{\sqrt{1+\left(1-\frac{r^6_H}{r^6}\right)(r\xi'_0)^2}}\right) =  k^2 r^5 \sin^3{2\xi_0} \sqrt{1+\left(1-\frac{r^6_H}{r^6}\right)(r\xi'_0)^2} \alpha_A \nonumber \\ & \quad +64 r \sin^3{2\xi_0} \sqrt{1+\left(1-\frac{r^6_H}{r^6}\right)(r\xi'_0)^2} \left[\frac{\ddot{\alpha_A}}{1-\frac{r^6_H}{r^6}}-\partial^2_{x_1} \alpha_A - \partial^2_{x_2} \alpha_A\right]  
      \eea
      
      \bea
      \partial_\theta (\sin{\theta} \partial_\theta \alpha_\theta)=\left(\frac{\cos{2\theta}}{\sin{\theta}}+k^2 \sin{\theta}\right)\alpha_\theta.
      \eea
      
As one can notice by inspection, $\alpha_\theta=C \sin{\theta}$ is a solution of the last equation, with $k^2=0$, obeying the boundary conditions. It is the lowest possible mode of $\alpha_\theta$, which is what we are after. One can explore higher modes but we will not do it here. With this, the first equation becomes:
      
      \bea
      & \partial_r \left[\frac{r^7 \sin^3{2\xi_0} \left(1-\frac{r^6_H}{r^6}\right) \alpha_A '}{\sqrt{1+\left(1-\frac{r^6_H}{r^6}\right)(r\xi'_0)^2}}\right] =  \nonumber \\ & \quad
       64 r \sin^3{2\xi_0} \sqrt{1+\left(1-\frac{r^6_H}{r^6}\right)(r\xi'_0)^2} \left[\frac{\ddot{\alpha_A}}{1-\frac{r^6_H}{r^6}}-\partial^2_{x_1} \alpha_A - \partial^2_{x_2} \alpha_A\right], 
      \eea
       which is identical to equation (\ref{eq:beta}). So, we find that $\alpha$ and $\beta$ have identical masses at least for the lowest $\alpha$ modes in $CP^3$. This can also be inferred from the symmetry of the embeddings. As previously mentioned, part of the symmetry of $RP^3$ is an $SO(3)$ symmetry which becomes a global symmetry in the field theory. The three scalars are a fundamental multiplet of this symmetry. In the massive embeddings this symmetry breaks to a $U(1)=SO(2)$ under which the fundamental multiplet breaks into a singlet and a two-vector. The fields $\beta, \alpha$ belong to the two-vector and therefore they still have the same mass, even to higher order. 
       
      Now we turn our attention to $\xi$. Expanding it as $\xi=\xi_0 + \delta\xi$ doesn't work. That is because fluctuations in the MN embedding need also to take into account slight movement in $r_0$, however the boundary conditions forces $\xi(r_0)=\frac{\pi}{2}, 0$ and therefore cannot allow such movement. So we need to convert to a different variable. We define a new radial coordinate $\rho$:
      
       \be
       (r_{H} \rho)^3 = r^3 \left(1+\sqrt{1-(\frac{r_{H}}{r})^6 }\right)
       \ee
       and new ``Cartesian" coordinates $u, y$: 
       
       \be
       u = \rho^2 \cos\left(2\xi-\frac{\pi}{2}\right)
       \ee
       
       \be
       y = \rho^2 \sin\left(2\xi-\frac{\pi}{2}\right).
       \ee
      
      We will describe the fluctuations in terms of $y(u)$\cite{Jensen}. Now we can expand the fields:
 
   \begin{figure}[h]
\center
\includegraphics[width=0.7\textwidth, height=0.35\textheight]{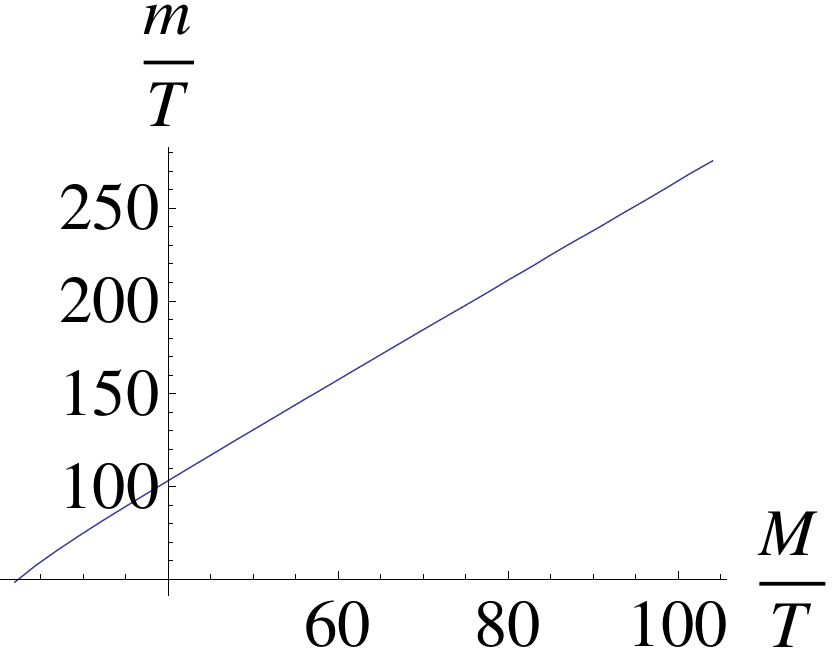}
\caption{The mass of the first mode of $y$ as a function of the embedding mass, for MN embeddings. Both masses are scaled by the temperature.}
\label{fig:fvsk}
\end{figure}

    \be
    y=y_0 + \delta y, \hspace{1cm} \frac{\theta_1-\theta_2}{2}=0, \hspace{1cm}  \frac{\phi_1+\phi_2}{2}=0.
    \ee 
      
      After evaluating the AdS determinant and the Chern-Simons term one gets the following Lagrangian:
      
      \bea
    &  L[\delta y] = L_0 + \Omega \frac{u}{\rho^2} \left[ \frac{ (1 - \rho^{12}) (\delta y')^2}{2 \rho^8 (\sqrt{1 + (y')^2})^3 } - \frac{ (1 - \rho^{12}) \tau \left(\dot{y}^2 \left(\frac{1 + \rho^6}{1 - \rho^6}\right)^2 -(\partial_{x_1} y)^2 - (\partial_{x_2} y)^2\right)}{2 \sqrt{1 + (y')^2} \rho^8 (1 + \rho^6)^{\frac{4}{3}}} \right. \nonumber \\ & \quad \left.  - \delta y' \delta y f(u,y,y')  -\frac{(\delta y)^2}{2} g(u,y,y')\right]
      \eea
      where
      
      \be
      f(u,y,y')=\frac{y y'}{\sqrt{1 + (y')^2}}\left(1+\frac{5}{\rho^{12}}\right) - \frac{1 - \rho^{12}}{3 \rho^{12}} \left[6 - u^2 \rho^{2}\left(1+\frac{7}{\rho^{6}}\right) \right]
      \ee
      and
      
      \bea
      g(u,y,y')=\sqrt{1 + (y')^2}\left[1 + \frac{5}{\rho^{12}} - \frac{y^2}{\rho^{4}} (1 + \frac{35}{\rho^{12}})\right] - \frac{6 y^3 (u y' - y)}{\rho^{8}}  \nonumber \\  + \frac{1 + \rho^{6}}{3\rho^{12}} \left[12 - 2 u^2 \rho^{2}\left(1+\frac{7}{\rho^{6}}\right) 
    + \frac{3y(u y' - y)}{\rho^{2}} \left(u^2 + 6 y^2 + \frac{7(u^2 - 2y^2)}{\rho^{12}}\right) \right].
      \eea
      
 $L_0$ is again the zeroth order Lagrangian giving the action (\ref{eq:xiL}) but recast in different variables, the tag stands for derivatives with respect to $u$, $\Omega$ is some numerical constant, and $\tau$ is given by:
      
      \be
      \tau=\frac{2^{\frac{16}{3}}}{r^4_{H}}.
      \ee
 There is also a first order term which vanishes by the EOM, as it must.
From this we get the EOM for $\delta y$:
      
      \bea
     & \partial_u \left(\frac{ u(1 - \rho^{12}) \delta y'}{\rho^{10} (\sqrt{1 + (y')^2})^3 }\right) -  \frac{ u (1 - \rho^{12}) \tau}{ \sqrt{1 + (y')^2} \rho^{10} (1 + \rho^6)^{\frac{4}{3}}}\left(\ddot{\delta y}(\frac{1 + \rho^6}{1 - \rho^6})^2-\partial^2_{x_1} \delta y-\partial^2_{x_2} \delta y\right) \\ \nonumber & \quad + \delta y \left( \frac{u g(u,y,y')}{\rho^{2}}-\partial_u(\frac{u f(u,y,y')}{\rho^{2}}) \right) =0
      \eea
      For a plane wave solution with $|\textbf{k}|=0$ we finally get:
      
      \bea
     & \partial_u \left(\frac{ u(1 - \rho^{12}) \delta y'}{\rho^{10} (\sqrt{1 + (y')^2})^3 }\right) + \frac{ u \tau \omega^2 \delta y  (1 + \rho^6)^{\frac{5}{3}}}{ \sqrt{1 + (y')^2} \rho^{10} (1 - \rho^6)} \\ \nonumber & \quad + \delta y \left( \frac{u g(u,y,y')}{\rho^{2}}-\partial_u(\frac{u f(u,y,y')}{\rho^{2}}) \right) = 0.
      \eea
    
    Expanding for large $u$ we now find exponential asymptotic behavior with one positive exponent, giving the non normalizable mode, and one negative giving the normalizable mode. 
      
    Next we solve the equation numerically for the lowest mode exactly as we did for $\alpha$ and $\beta$. The result is shown in figure 6.

In BH embeddings, again there are no normalizable modes. This follows from the equation, as expanding it around the horizon ($\rho^4 = 1 + \epsilon$) we find that:

\be
 y\approx A\cos{\left(\frac{8\omega}{3 r^2_H}\ln{\epsilon}\right)} + B\sin{\left(\frac{8\omega}{3 r^2_H}\ln{\epsilon}\right)}.
\ee
        
    \section{Conductivities}
    
    Next we will calculate the conductivities. We will do this in a more general case in which there is a $B_2$ field (\ref{eq:bt}). This system is interesting as it might be used as a holographic model for the QHE. The reason is that in this model we have the following CS term on the D6-brane:
    
    \be
      \int C_1 \wedge B_2  \wedge F \wedge F \rightarrow \int_{AdS_4} \theta F \wedge F. \label{eq:CS}
    \ee
     This seems to describe a 3d system of gauge and matter fields with a CS term and therefore we expect QHE conductivities analogous to\cite{BJLL}. However this system is supersymmetric and it seems strange that it will have a term such as (\ref{eq:CS}). The field theory corresponds to the 3d intersection of the D6-brane with various branes preserving ${\cal N}=3$ supersymmetry, and so should not generate any CS term. This is also apparent from the quiver diagram (Fig. 1). Since the added flavors are a hypermultiplet in the fundamental representation, the contributions to the CS term from the fermions should cancel. In order to better understand this we will calculate the conductivities. 
    
    Following \cite{KO}, we will add a non-constant gauge field on the probe brane:
    
    \be
    A_i = \frac{R^3 }{4\pi k} \left( t e_i + \frac{1}{2}\epsilon_{ij} x_j b + a_i (r) \right)
    \ee
    
    \be
    A_0 = \frac{R^3 }{4\pi k} a_0 (r),
    \ee    
    where $e_i, b$ are the externally applied electric and magnetic fields respectively, and $a_i, a_0$ will be given by the equation of motion and they will determine the currents. All the fields have been conveniently scaled.
 
 In order to evaluate the conductivities we need to calculate the action. We will start with the DBI part, but before that we need to point out that there is also a CS term of the form:
 
    \be
     S = 2\pi \int C_3 \wedge B_2 \wedge F
    \ee
   that sources the $CP^3$ components of the gauge field for the background of interest. So we must also include the terms:
  
  \be
  A_{\psi} = \frac{R^3 }{4\pi k} a_{\psi} (r,\theta)  \hspace{1cm}   A_{\phi} = \frac{R^3 }{4\pi k} a_{\phi} (r,\theta).
  \ee 
    To simplify matters, I will work with the following ansatz:
    
    \be
  a_{\psi} =  a_{\psi} (r)  \hspace{1cm}   a_{\phi} =  (a_{\psi} (r) + C_{\psi}) \cos(\theta).
  \ee  
  This solves the $a_{\phi}$ EOM, and furthermore it is necessary if one wants $\xi, a_0$ and $a_i$ to be $\theta$ independent. $C_{\psi}$ is an arbitrary constant that for simplicity I will set to zero. Now I can evaluate the DBI part, which after a lengthly calculation gives:
    
    \bea
     S_{DBI} =  \frac{R^{9}}{2^{7}k^{2}} \int dr \sin{2\xi} \sqrt{1+(a_{\psi}+\ell \cos{2\xi})^2} \sqrt{ f(r)},  
    \eea
    where
    
    \bea
   & f(r)= 4\left(\frac{1}{r^2 (1-\frac{r^6_H}{r^6})}+\left((\xi')^2 + \frac{1}{4}\left(\frac{a'_{\psi}}{\sin{2\xi}}-2 \ell \xi'\right)^2\right)\right)(\frac{r}{2})^4 \beta(r) \\ \nonumber & \quad -(b a'_0 + \vec{a}'\times\vec{e})^2 + (\frac{r}{2})^8 \left( (1-\frac{r^6_H}{r^6})|\vec{a}'|^2-(a'_0)^2\right)
    \eea
    and where we have defined:
   
   \bea
    \beta(r) \equiv (1-\frac{r^6_H}{r^6})(b^2 + (\frac{r}{2})^8)-|e|^2
    \eea
    and
    
    \be
     \ell \equiv \frac{l}{R^3}.
    \ee 
     
    The relevant CS terms are:
    
    \bea
     S_{CS} & = & \frac{1}{2} \int C_3 \wedge (B_2 + 2 \pi F) \wedge (B_2 + 2 \pi F) \\ \nonumber & + & (2\pi)^2  \int C_1 \wedge (B_2 + 2 \pi F) \wedge (B_2 + 2 \pi F) \wedge (B_2 + 2 \pi F).
    \eea
   Combining all of these together we get the action:
    
    \bea
    S & = & \int dr  \left[\frac{R^{9}}{2^{7}k^{2}} \sin{2\xi} \sqrt{1+(a_{\psi}+\ell \cos{2\xi})^2} \sqrt{ f(r)} - \frac{R^{9}r^{6}}{3\cdot2^{13}k^{2}} \sin 4\xi  \xi' \right. \nonumber \\  &-& \frac{R^{9} r^6}{2^{10} k^2} \left(\ell^2 \sin(4\xi)\xi' -a_{\psi}(a'_{\psi}-2\ell \sin(2\xi)\xi')-\ell \cos(2\xi)a'_{\psi}\right) \label{eq:scon} \\ \nonumber    &+& \left. \frac{R^9}{16 k^2} \cos(2\xi) (\vec{e}\times\vec{a'}-b a'_0) (a_{\psi}+\ell \cos(2\xi))\right]. 
    \eea
    
    As $S$ depends on $a_0, a_i$ only through derivatives, there are 3 conserved quantities:   
       
    \be
      \rho = \frac{\partial L}{\partial a'_0}
    \ee
    
    \be
      j_{||} = \frac{16k^2}{R^9} \frac{\partial L}{\partial a'_{||}}
    \ee
    
    \be
      j_{\bot} = \frac{16k^2}{R^9} \frac{\partial L}{\partial a'_{\bot}},
    \ee
   where $j_{\bot}, j_{||}$ are the components transverse and parallel to the electric field, respectively. It can be shown \cite{KO} that these are the physical charge and current densities. From this we can, after a prolonged calculation, get the following relations:
    
    \bea
      a'_0(r) & = & \frac{8}{r^3}\sqrt{1+(1-\frac{r^6_H}{r^6})\left((r\xi')^2+\frac{r^2}{4}(\frac{a'_{\psi}}{\sin{2\xi}}-2\ell\xi')^2\right)} \nonumber \\ & \times & \frac{\left(\tilde{\rho} \left(|e|^2 - (\frac{r}{2})^8 (1-\frac{r^6_H}{r^6})\right) - b |e| \tilde{j_{\bot}} \right)}{\sqrt{\beta(r)\gamma(r)-\alpha^2(r)}}\label{eq:az}
    \eea
    
    \bea
      a'_{\bot}(r) =  a'_0 \frac{\tilde{j_{\bot}}(b^2 + (\frac{r}{2})^8)-|e| b \tilde{\rho}}{\tilde{\rho} \left(|e|^2 - (\frac{r}{2})^8 (1-\frac{r^6_H}{r^6})\right) - b |e| \tilde{j_{\bot}}}
    \eea
    
   \bea
    a'_{||}(r) =  a'_0 \frac{j_{||} (b^2 + (\frac{r}{2})^8 - \frac{|e|^2}{1-\frac{r^6_H}{r^6}})}{\tilde{\rho} \left(|e|^2 - (\frac{r}{2})^8 (1-\frac{r^6_H}{r^6})\right) - b |e| \tilde{j_{\bot}} }, \label{eq:ai}
    \eea
    where:
    
    \be
    \tilde{\rho}(r)\equiv \rho - b\cos{2\xi}(a_{\psi}+\ell\cos{2\xi}) 
    \ee
    
    \be
    \tilde{j_{\bot}}(r)\equiv j_{\bot} - |e| \cos{2\xi}(a_{\psi}+\ell\cos{2\xi})
    \ee
       
    \bea
    \gamma(r) &\equiv& (1-\frac{r^6_H}{r^6})\left(\tilde{\rho}^2 + (\frac{r}{2})^8 R^6 \sin^2{2\xi}\left(1+4(a_{\psi}+\ell \cos{2\xi})^2\right)\right) \nonumber \\ &-& \frac{(\vec{\tilde{j}} \times \vec{e})^2 +(\vec{j} \cdot \vec{e})^2}{|e|^2}
    \eea
    
    \bea
    \alpha(r) \equiv \vec{\tilde{j}} \times \vec{e}-b\tilde{\rho} (1-\frac{r^6_H}{r^6}).
    \eea

    Inserting this in the DBI part we get:
    
    \bea
    & S_{DBI} & =  \frac{R^{12}}{2^{4}k^{2}} \int dr    \frac{\sin^2(2\xi) Y(r) \label{eq:sf} \beta(r)}{\sqrt{\beta(r)\gamma(r)-\alpha^2(r)}}  \\ \nonumber & \times &   \sqrt{1+(1-\frac{r^6_H}{r^6})((r\xi')^2+\frac{r^2}{4}(\frac{a'_{\psi}}{\sin{2\xi}}-2\ell\xi')^2)}, 
    \eea
    where:
    
    \be
     Y(r) = (\frac{r}{2})^3 \sqrt{1-\frac{r^6_H}{r^6}} \sqrt{1+4(a_{\psi}+\ell \cos{2\xi})^2}\sqrt{1+(a_{\psi}+\ell \cos{2\xi})^2}.
    \ee
    
    \subsection{BH embeddings}
    
    For BH embeddings the calculation follows as in\cite{KO,O}. The argument in the square root in the denominator of (\ref{eq:sf}) must not be negative, however the positive factor is zero when:
    
    \bea
     \beta(r)=0 \rightarrow |e|^2 = (1-\frac{r^6_H}{r^6})(b^2 + (\frac{r}{2})^8)
    \eea
    which always has a solution for some $r_* > r_H$.
    The only way the square root can be non-negative there is if:
    
    \bea
     \alpha(r_*)=0 \rightarrow \vec{\tilde{j}}_* \times \vec{e} - b\tilde{\rho}_* (1-\frac{r^6_H}{r^6_*}) = 0
    \eea
    where $\tilde{\rho}_* = \tilde{\rho}(r_*)$ and $\tilde{j}_* = \tilde{j}(r_*)$.
    Furthermore, when $r<r_*$, $\beta$ becomes negative, and the only way the entire square root can stay positive is if $\gamma$ also becomes negative in that range. Therefore, that factor must also vanish at that point, so in addition:
    
    \bea
     \gamma(r_*) &=& 0 \rightarrow  (1-\frac{r^6_H}{r^6_*})^{-1}\left((\tilde{j_{* \bot}})^2 + (j_{* ||})^2\right) \nonumber \\  &=&  \tilde{\rho_*}^2 + (\frac{r_*}{2})^8 R^6 \sin^2{2\xi_*}\left(1+4(a_{* \psi}+\ell \cos{2\xi_*})^2\right).
    \eea
    
    From this we can derive a relation between the currents and the external fields and thus calculate the conductivity. When the dust settles we get:
    
    \bea
    \sigma_{\bot} = \frac{b\tilde{\rho_*}}{b^2 + (\frac{r_*}{2})^8}
    \eea
    
    \bea
    \sigma_{||} = \sqrt{\frac{\tilde{\rho_*}^2 + (\frac{r_*}{2})^8 R^6 \sin^2{2\xi_*}(1+4(a_{\psi *}+\ell \cos{2\xi_*})^2)}{b^2 + (\frac{r_*}{2})^8}-\frac{b^2 \tilde{\rho_*}^2}{(b^2 + (\frac{r_*}{2})^8)^2}}
    \eea
    The conductivity in the BH embeddings is similar to that of a normal metal. In particular the transverse conductivity depends linearly on $\rho$ and $\frac{1}{b}$ (at least when $b>>r^4_H$).
    
    \subsection{MN embeddings}
    
    The previous calculation is true only in the BH embeddings, as we must assume that the brane fills the all range $r_H<r<\infty$, otherwise it may end before $r_*$ and the argument breaks down. However, in the MN embeddings case we have a boundary condition coming from finiteness of the derivatives of the components of the gauge field. This is necessary as divergences in the derivatives of the gauge field signal that there are sources at the tip. This means that there are strings connecting the probe brane with the base branes. As it turns out, these strings exert a force on the probe brane forcing it to turn to a BH embedding\cite{KMMMM}. So in order to have MN embeddings we must have no sources, which therefore also requires that the derivatives of the components of the gauge field be finite. As we previously mentioned, $\xi'$ diverges at $r=r_0$. At that point $a'$ must be finite. Equations (\ref{eq:az}-\ref{eq:ai}) reveal that the only way that these terms can be finite is if:
    
    \be
    \tilde{\rho}(r_0)=0  \rightarrow \rho=b\cos{2\xi}(a_{\psi}+\ell\cos{2\xi})|_{\xi\rightarrow\frac{\pi}{2}} = b(\ell - a_{\psi}(r_0))
    \ee
    
    \be
      j_{||}=0 \rightarrow \sigma_{||} = 0
    \ee  
       
    \be
    \tilde{j_{\bot}}(r_0) =0 \rightarrow \sigma_{\bot} = \cos{2\xi}(a_{\psi}+\ell\cos{2\xi})|_{\xi\rightarrow\frac{\pi}{2}} = \ell - a_{\psi}(r_0).
    \ee   
          
          We see that in the MN case we get a vanishing longitudinal conductivity and a non-zero transverse conductivity. As MN embeddings possess a mass gap we expect them to be insulators, therefore the vanishing of $\sigma_{||}$ is expected. The non-zero transverse conductivity is surprising and suggests that the system is in a quantum hall state. However that is inconsistent with the supersymmetry. 
          
          In order to better understand this we shall look at the $a_{\psi}$ EOM, which is\footnote{It is important to note that the equation cannot be derived from the action (\ref{eq:scon}), but must be derived from the more general Lagrangian, applying the ansatz afterward.}: 
          
          \bea
         & \partial_{r} \left( \frac{r^4 \beta(r)\left( a'_{\psi} - 2\ell \sin(2\xi)\xi' \right)\sqrt{1+(a_{\psi}+\ell \cos{2\xi})^2}}{\sin(2\xi)\sqrt{f(r)}} \right) +\frac{3r^5}{4}(a_{\psi}+\ell \cos{2\xi}) = \\ \nonumber & \quad \frac{(a_{\psi}+\ell \cos{2\xi})\sin(2\xi)\sqrt{f(r)}}{\sqrt{1+(a_{\psi}+\ell \cos{2\xi})^2}}+16\cos(2\xi)\left( b a'_0 + \vec{a'}\times\vec{e} \right).
          \eea
          
          First, one can notice that there is the following solution to all the gauge field EOM:
          
          \be
          a'_0 = \vec{a'} = a_{\psi}+\ell \cos{2\xi} =0, \label{eq:nc}
          \ee
          which implies that $\sigma_{\bot} = 0$.
          This solution corresponds to $2\pi F = -B_2$, and sets the charge density and currents to zero regardless of $e$ and $b$. Furthermore one can also find MN solutions to the $\xi$ EOM. This solution appears to be consistent with what we know about the field theory on the D6-brane. 
          
          However, there may be other solutions with a non-zero value of $a_{\psi}+\ell \cos{2\xi}$ at the tip, and therefore non-zero currents. These will generate flux on the two-sphere spanned by the coordinates $\theta, \phi$, given by:
          
          \bea
          2\pi F_{\theta \phi} + B_{2,\theta \phi}= \partial_{\theta} a_{\phi}-\frac{l}{2k}\sin(\theta)\cos(2\xi) \nonumber \\  = -\frac{R^3}{2k}\sin(\theta) \left(a_{\psi}+ \ell \cos(2\xi)\right).   
          \eea
          This is important since this two-sphere shrinks to zero at the tip so the total flux on it must be zero. Again, this is required by the absence of sources, which whould render the MN embedding unstable. This then forces $a_{\psi}+\ell \cos{2\xi} =0$ to be zero everywhere, and thus leads to a vanishing transverse conductivity. 
          
    \section{Conclusion}
    
    We analyzed the ABJM model with fundamental massive matter and showed that it exhibits a first order phase transition between solutions which end before the horizon (favored at low temperature) and those that enter the black hole (favored at high temperature). This is similar to the D3/D7 system which exhibits the same phenomenon. In the field theory this is a transition between an insulator (MN embedding), and a metallic state (BH embedding). 
    
    We have shown that the conductivities of the BH embedding are as expected in a metallic state. We have also shown that the longitudinal conductivity vanishes in the MN embedding as expected. We have further argued that the transverse conductivity must also vanish, in accordance with the supersymmetry.
    
\subsection*{Acknowledgments}

    I thank Oren Bergman, Niko Jokela and Gilad Lifschytz for useful comments and discussions. G.Z. is supported
in part by the Israel Science Foundation under grant no. 392/09.


\begin{thebibliography}{40}

\bibitem{ABJM}
  O.~Aharony, O.~Bergman, D.~L.~Jafferis and J.~Maldacena,
  JHEP {\bf 0810}, 091 (2008)
  [arXiv:0806.1218 [hep-th]].
  
\bibitem{HK}
  S.~Hohenegger, I.~Kirsch,
  JHEP {\bf 0904}, 129 (2009)
  [arXiv:0903.1730 [hep-th]].
  
\bibitem{GJ}
  D.~Gaiotto, D.~L.~Jafferis,
  [arXiv:0903.2175 [hep-th]].
  
\bibitem{HLT}
  Y.~Hikida, W.~Li, T.~Takayanagi,
  JHEP {\bf 0907}, 065 (2009)
  [arXiv:0903.2194 [hep-th]].
  
\bibitem{AEMOT}  
  M.~Ammon, J.~Erdmenger, R.~Meyer, A.~O'Bannon and T.~Wrase,
  JHEP {\bf 0911}, 125 (2009)
  [arXiv:0909.3845 [hep-th]].

\bibitem{FT}  
  M.~Fujita, T.~S.~Tai, 
  JHEP {\bf 0909}, 062 (2009)
  [arXiv:0906.0253 [hep-th]].
  
\bibitem{Jensen}
  K.~Jensen, 
  [arXiv:1006.3066 [hep-th]].
 
\bibitem{BEEZI}  
  J.~Babington, J.~Erdmenger, N.~Evans, Z.~Guralnik and I.~Kirsch,
  Phys.\ Rev.\  D {\bf 69}, 066007 (2004)
  [arXiv:0306018 [hep-th]].
  
\bibitem{GSUM}  
  K.~Ghoroku, T.~Sakaguchi, N.~Uekusa and M.~Yahiro,
  Phys.\ Rev.\  D {\bf 71}, 106002 (2005)
  [arXiv:0502088 [hep-th]].
  
\bibitem{MMT}  
  D.~Mateos, R.~C.~Myers, R.~M.~Thomson 
  JHEP {\bf 0705}, 067 (2007)
  [arXiv:0701132 [hep-th]].

\bibitem{ABJ}
  O.~Aharony, O.~Bergman, D.~L.~Jafferis,
  JHEP {\bf 0811}, 043 (2008)
  [arXiv:0807.4924 [hep-th]].

\bibitem{BJLL}
  O.~Bergman, N.~Jokela, G.~Lifschytz and M.~Lippert,
  [arXiv:1003.4965 [hep-th]].
     
\bibitem{KO}
  A.~Karch, A.~O'Bannon, 
  JHEP {\bf 0709}, 024 (2007)
  [arXiv:0705.3870 [hep-th]].

\bibitem{O}
  A.~O'Bannon, 
  Phys.\ Rev.\  D {\bf 76}, 086007 (2007)
  [arXiv:0708.1994 [hep-th]].
  
\bibitem{KMMMM}  
  S.~Kobayashi, D.~Mateos, S.~Matsuura, R.~C.~Myers and R.~M.~Thomson 
  JHEP {\bf 0702}, 016 (2007)
  [arXiv:0611099 [hep-th]].

\end{thebibliography}
\end{document}